\documentstyle[12pt]{JHEP}

\font\blackboard=msbm10 at 12pt
\font\blackboards=msbm7
\font\blackboardss=msbm5
\newfam\black
\textfont\black=\blackboard
\scriptfont\black=\blackboards
\scriptscriptfont\black=\blackboardss
\def\bb#1{{\fam\black\relax#1}}

\def\br{\bb R}

\newcommand{\NP}{{\em Nucl.\ Phys.\ }}

\newcommand{\PL}{{\em Phys.\ Lett.\ }}

\newcommand{\MPL}{{\em Mod.\ Phys.\ Lett.\ }}

\newcommand{\gone}[1]{}

\title{D2-branes in B fields}
\author{Washington Taylor\\
{Center for Theoretical Physics} \\
{MIT, Bldg.  6-306} \\
{Cambridge, MA 02139, U.S.A.} \\
{\tt wati@mit.edu}}

\abstract{This note focuses on the coupling of a type IIA D2-brane to
a background B field.  It is shown that the D0-brane charge arising
from the integral over the D2-brane of the pullback of the B field is
cancelled by bulk contributions, for a compact D2-brane wrapping a
homotopically trivial cycle in space-time.  In M-theory this
cancellation is a straightforward consequence of momentum
conservation.  This result resolves a puzzle recently posed by Bachas,
Douglas and Schweigert related to the quantization of R-R charges on
stable spherical D2-branes on the group manifold SU(2).}

\keywords{D-branes, M-theory}
\preprint{MIT-CTP-2973, hep-th/0004141}
\begin{document}

\section{Introduction}

A D2-brane in a general type IIA supergravity background has a
world-volume action given by a sum of Born-Infeld and Wess-Zumino
terms \cite{Polchinski-TASI,cgnsw,Bergshoeff-Townsend}
\[
S = S_{{\rm BI}} + S_{{\rm WZ}}
\]
The Born-Infeld part of the action is given by \cite{Leigh}
\[
S_{{\rm BI}} = -T_2 \int e^{-\Phi} \sqrt{-\det (G_{\alpha \beta} +{\cal
F}_{\alpha \beta})} 
\]
where $G_{\alpha \beta}$ is the pullback of the space-time metric to
the D2-brane world-volume, $\Phi$ is the pullback of the dilaton and
and
\[
{\cal F}_{\alpha \beta} = 2 \pi \alpha' F_{\alpha \beta}
-B_{\alpha \beta} 
\]
combines the pullback of the space-time B field and the field strength
$F$ of the U(1) gauge field living on the brane.

The Wess-Zumino terms couple the brane to the space-time R-R fields
through \cite{Li-bound,Douglas}
\[
S_{{\rm WZ}}= \int \left(\sum_{i}^{} C^{(i)} \right) \wedge e^{\cal F}.
\]
In particular there is a term in the D2-brane action of the form
\begin{equation}
\int C^{(1)} \wedge ( 2 \pi \alpha' F -B).
\label{eq:term}
\end{equation}
This is the term which we will discuss in this note.  In particular,
we will focus on the coupling between the space-time R-R 1-form field
$C^{(1)}$ and the B field which is mediated by the D2-brane in this
interaction term.

For a compact D2-brane of arbitrary topology, the first Chern class of
the U(1) bundle on the brane at a fixed point in time gives an integer
\[
\frac{1}{2 \pi}  \int F  = k.
\]
According to (\ref{eq:term}), this integer acts as a source for the
R-R vector field and represents $k$ D0-branes bound to the D2-brane
\cite{Douglas}.  This interpretation of the U(1) flux on the D2-brane
also naturally follows from T-duality \cite{Bachas,WT-Trieste}.  The
quantization of $\int F$ corresponds nicely with our expectation that
the number of D-particles in any system is integral.

The coupling of the R-R vector field to the pullback of the space-time
B field on the D2-brane is at first sight somewhat more surprising, as
there is no natural reason for the integral $\int B$ to be quantized.
Indeed, in a region of constant $H = dB$ flux one can imagine blowing
up a small spherical D2-brane out of the vacuum, which would have a
continuously varying value of $\int B$.  This lack of quantization of
$\int B$ was found by Bachas, Douglas and Schweigert to be
particularly puzzling in the context of D-branes on the $SU(2)$ group
manifold, where stable spherical D2-branes are predicted by conformal
field theory \cite{bds}.  These stable spherical D2-branes would seem
from (\ref{eq:term}) to have non-integral, and in fact irrational
values of D0-brane charge.

In this note we clear up this puzzle.  We show that for a compact
D2-brane embedded on a homotopically trivial spatial cycle the
D0-brane charge associated with $\int B$ is precisely cancelled by a
contribution from the bulk fields.  This cancellation is guaranteed by
conservation of D0-brane charge in the full supergravity + D2-brane
theory.  From the point of view of M-theory, this cancellation can be
seen in terms of conservation of momentum in the compact direction.

In Section 2 we give a brief discussion of an analogous situation in
4D classical electromagnetism which should make the physical argument
quite transparent.  In Section 3 we repeat this analysis in the
context of M-theory, and in Section 4 we interpret the discussion in
terms of the IIA language.  Section 5 contains a discussion of the
connection with other work and the resolution of the puzzle posed by
Bachas, Douglas and Schweigert.  After this note was written, we
learned that a similar resolution of this puzzle has been found by
Polchinski \cite{Polchinski-private}.

\section{Electromagnetic analogy}

Consider classical electromagnetism in 4 dimensions.  Let us restrict
attention to physical configurations which are invariant under
translation in the $X^{3}$ direction, so that we may choose a gauge in
which the vector potential $A_\mu$ is independent of this coordinate.
Let us assume that there is a background magnetic field $F_{i3} =
\partial_i A_3$ present, for $i = 1, 2$.

Now, let us separate a pair of equal and opposite charges $+ q, -q$ by
starting with both charges at a point ${\bf a}$ and moving the charge
$+ q$ along a path ${\cal P}$ in the 1-2 plane to the point ${\bf b}$.
This gives us an electric dipole.  In the process of separating the
charges we move the positive charge through the magnetic field
$F_{i3}$.  The Lorentz force ${\bf F} = q {\bf v} \times {\bf B}$
gives a net momentum in the $X^3$ direction
\begin{eqnarray}
p_3 & = & - q\int_{{\rm a}}^{ {\rm b}} F_{i3}  \;dx^i
\label{eq:momentum}\\
 & = & q \left(A_3 ({\bf a})-A_3 ({\bf b})\right) \nonumber
\end{eqnarray}
to the electric dipole.  
Note that this momentum is independent of the path ${\cal P}$ since
$dF = 0$.
By conservation of momentum, this momentum
must be cancelled by the momentum in the electromagnetic fields.
Indeed, we can compute the momentum in the Poynting vector flux ${\bf
E} \times {\bf B}$, of which the $X^3$ component is
\begin{eqnarray*}
\int F_{0i} F^{i3}  & = & \int F_{0i} \;\partial_iA_3\\
 & = & -\int A_3 \; \partial_iF_{0i}\\
& = & \int A_3 \;q (\delta ({\bf x}- {\bf b})-\delta ({\bf x} -
{\bf a}))\\
& = & q \left(A_3 ( {\bf b}) -A_3 ({\bf a}) \right)
\end{eqnarray*}
which precisely cancels (\ref{eq:momentum}).

This calculation is easily generalized to show that any static
configuration of charges with net charge 0 in the presence of a
magnetic field $F_{i 3}$ has a total momentum contained in the
electromagnetic fields whose component in the $X^3$ direction is equal
and opposite to the momentum of the charges when the charges each are
taken to have total momentum
\[
p_3 = -q A_3.
\]
This calculation is precisely analogous to the result for a membrane
moving in a background  4-form field strength $F_{\mu \nu \lambda \, 11}$
in M-theory, which we now discuss.

\section{M-theory picture}

The low-energy description of M-theory is given by 11-dimensional
supergravity.  In addition to the metric tensor and gravitino field,
11D supergravity has a dynamical 3-form field $C_{IJK}$ which is
closely analogous to the U(1) vector field $A_\mu$ of 4D
electromagnetism.  M-theory contains dynamical membranes which couple
electrically to the 3-form field through a term of the form
\[
\int_V d^3 \xi^\alpha \; C
\]
where $V$ is the membrane world-volume and $C$ is the pullback of the
3-form to $V$.  The curvature of $C$ is a 4-form $F = dC$ with components
\[
 F_{IJKL} = 4\,\partial_{[I} C_{JKL]}.
\]

Let us consider a compact membrane $\Sigma$ of arbitrary genus
embedded in a space of topology $\br^{10}$, in the presence of a
magnetic field strength $F_{\mu \nu \lambda \, 11}$.  Although there
may be forces such as the membrane tension
acting on the membrane we can imagine that the membrane
is held in a static position by some additional external forces.  With
the application to type IIA D2-branes in mind, we will imagine that
the field configuration is independent of $X^{11}$ so that all
components of $C_{IJK}$ can be chosen to be independent of $X^{11}$.
Since our membrane is homotopically trivial, we can imagine starting
with the entire membrane at a fixed point $a$ in space and expanding
the membrane to its desired shape.  We can
parameterize this family of deformations of the membrane with a
parameter $\tau \in [0, 1]$.  We denote by $\Gamma$ the 3-volume swept
out by $\Sigma\times[0, 1]$ as we vary $\tau$.  In performing the
expansion of the membrane from a point to the desired geometry
$\hat{\Sigma} = \partial \Gamma$, we must move the membrane in the
background magnetic field, which imparts to it a net momentum in the
$X^{11}$ direction
\begin{eqnarray}
p_{11} & = &-\frac{1}{6} 
\int_\Gamma F_{ijk\, 11}  \;d X^i \wedge d
X^j\wedge d X^k
\label{eq:M-momentum}\\
 & = &  -\frac{1}{2}
\int_{\hat{\Sigma}} C_{ij\, 11}\;
d X^i \wedge d X^j \nonumber.
\end{eqnarray}
This is the analogue in M-theory of the momentum arising from the
force $q {\bf v} \times {\bf B}$ in classical electromagnetism.  Note
that this quantity is invariant under ($X^{11}$-independent) gauge
transformations of the form $\delta C_{ij\,11} = \partial_i
\Lambda_j-\partial_j \Lambda_i$ when the membrane is wrapped on a
homotopically trivial cycle since $dd \Lambda = 0$.  

Because momentum in M-theory is conserved, the momentum imparted to
the membrane in this process must be balanced by a momentum flux in
the 4-form field strength.  Indeed, there is a contribution to the
11-momentum in M-theory from the term in the stress tensor
\[
\frac{1}{6} \,F_{0ijk} F^{ijk\,11}
\]
Integrating by parts, we can rewrite this contribution to the momentum
just as in the electromagnetic analogue by
\begin{eqnarray}
\frac{1}{6}
\int F_{0ijk} F^{ijk\,11} & = &  \int F_{0ijk} \; \partial_{[i}C_{jk]\,
 11} \label{eq:m-Poynting}\\
 & = &  -\int C_{[jk\,
 11}\; \partial_{i]} F_{0ijk}\nonumber\\
& = & \frac{1}{2}\int_{\hat{\Sigma}} C_{ij\, 11}\;
d X^i \wedge d X^j \nonumber.
\end{eqnarray}
This precisely cancels the momentum of the membrane (\ref{eq:M-momentum})
given by its
motion in the magnetic field.

Thus, we see that it is natural to associate with an M-theory membrane
in a magnetic field an intrinsic momentum
\begin{equation}
-\frac{1}{2} \int_{\Sigma} C_{ij\, 11}\;
d X^i  \wedge d X^j \nonumber
\label{eq:membrane-momentum}
\end{equation}
which is precisely cancelled by the momentum in the 4-form field
produced by the interaction between the ``electric'' field $F_{0ijk}$
produced by the membrane and the external ``magnetic'' field $F_{jkl\,
11}$ in which the membrane is sitting.

\section{IIA picture}

We would now like to translate the preceding discussion into the
language of type IIA string theory.  When 11-dimensional supergravity
is dimensionally reduced by compactifying on a small circle in the
$X^{11}$ direction, the resulting theory is type IIA supergravity.
Under this dimensional reduction, the 3-form field $C_{IJK}$ decomposes
into the R-R 3-form field $C^{(3)}_{\mu \nu \lambda}$ and the NS-NS
2-form field $B_{\mu \nu}$ of the IIA theory.  The R-R 1-form field
$C^{(1)}_{\mu}$ in the IIA theory arises from the Kaluza-Klein vector
field $g_{\mu \; 11}$, and the quanta of momentum in the compact
direction of M-theory are then associated with D0-branes, which are
the objects carrying charges under $C^{(1)}$.

With these identifications, we see that the 11-momentum associated
with a membrane in a background magnetic field
(\ref{eq:membrane-momentum}) becomes a D0-brane charge on a D2-brane
$\Sigma$ associated with the integrated pullback of the B field
\begin{equation}
-\int_\Sigma B.
\label{eq:b-charge}
\end{equation}
By the argument described above, this D0-brane charge must be
cancelled by an additional contribution to the D0-brane charge
associated with fields in the bulk.  Indeed, just such a term appears
in the action of type IIA supergravity.  The curvatures of the IIA fields
$C^{(3)}$ and $B$  are given by
\begin{eqnarray}
H & = &  dB \nonumber\\
G^{(4)} & = &  dC^{(3)} + C^{(1)} \wedge H.  \label{eq:g-definition}
\end{eqnarray}
In the IIA supergravity action there is a term quadratic in the
curvature $G^{(4)}$
\[
-\frac{1}{48}  \int d^{10} x \; \sqrt{-g} | G^{(4)} |^2.
\]
From the definition (\ref{eq:g-definition}), we see that this includes
a term proportional to
\begin{equation}
(C^{(1)} \wedge dB) \cdot  (d C^{(3)}).
\label{eq:IIA-Poynting}
\end{equation}
This is just the term we need to cancel (\ref{eq:b-charge}), just as
the bulk Poynting-type contribution to the 11-momentum
(\ref{eq:m-Poynting}) cancels the membrane momentum
(\ref{eq:membrane-momentum}).  Indeed, the term contributing to IIA
D0-brane charge in (\ref{eq:IIA-Poynting}) is precisely the
dimensional reduction of (\ref{eq:m-Poynting}).

To complete the story we can simply check that the D0-brane charge
contained in (\ref{eq:IIA-Poynting})
indeed can be related through integration by parts to the negative of
the D0-brane charge (\ref{eq:b-charge})
\begin{eqnarray}
\frac{1}{6}\int G^{(4)}_{0ijk} H^{ijk} & = & 
\int G^{(4)}_{0ijk} \partial_{[i}B_{jk]} \label{eq:IIA-integration}\\
 & = & - \int B_{[jk} \partial_{i]} G^{(4)}_{0ijk}\nonumber\\
& = & \int_\Sigma B.\nonumber
\end{eqnarray}

Thus, we have shown that the D0-brane charge associated with a
D2-brane in an external B field is cancelled by a contribution from
the bulk when the D2-brane is embedded on a homotopically trivial
cycle.  Indeed, by directly using the integration by parts argument in
(\ref{eq:IIA-integration}), we see that this result holds whenever
there is no boundary contribution to the integral of the bulk
contribution to the D0-brane charge.  Note, however, that the
interpretation of (\ref{eq:b-charge}) as being the total charge
arising from the analogue of the Lorentz force is only valid when the
D2-brane can be homotopically contracted to a point.

\section{Discussion and examples}

We have shown that the term $\int C^{(1)} \wedge B$ in the
world-volume action of a D2-brane should be associated with a D0-brane
charge which is generally cancelled by an opposite contribution from
the bulk fields.  In M-theory, this cancellation is a simple
consequence of momentum conservation.

There are several situations in which this interpretation of the B
field contribution to D0-brane charge on a membrane is useful.  For
one thing, the cancellation of this contribution to D0-brane charge
clarifies the question of quantization of D-particle number in a type
IIA configuration containing D2-branes and B field fluxes.  Because
the contribution from $\int B$ to the D0-brane charge is always
cancelled in the bulk, the quantization of D-particle number is
automatically guaranteed by the topological condition that the
integral of the U(1) flux $\int F$ is quantized in units of $2 \pi$.
In particular, this clears up the puzzle posed by Bachas, Douglas and
Schweigert in \cite{bds}.  They found a set of stable spherical
D2-branes on the group manifold $SU(2)$, with integrated B fields
which seemed to indicate irrational values for D0-brane charge.  From
the discussion in this note, it is clear that these B field
contributions to the D0-brane charge are cancelled by Poynting-type
bulk contributions of the form (\ref{eq:IIA-integration}).  Indeed,
one could imagine adiabatically moving between any pair of the stable
spherical D2-branes found in \cite{bds}.  In this process, the
D2-brane would pick up additional 11-momentum (D-particle charge) from
the analogue of the Lorentz force condition, which would be signified
by the change in $\int B$.  At the same time, the D2-brane would act
back on the fields, increasing the net D-particle charge in the bulk
from the analogue of the Poynting flux.  Thus, we see that the results
of \cite{bds} are perfectly consistent and that there is no paradox:
D0-brane charge is always integrally quantized, and only arises from
free D-particles or from the $p$th Chern class of the U(1) field on a
D2$p$-brane.

Another situation in which the results in this note are relevant is
when a D2-brane bubble is produced from a system of $N$ D0-branes
through the introduction of a background electric 4-form field.  It
was shown in \cite{Mark-Wati-4} that there is a term in the action
describing a system of multiple D0-branes in a background 4-form field
of the form
\begin{equation}
{\rm Tr}\;\left([X^i, X^j] X^k \right)
\; G^{(4)}_{0ijk}.
\label{eq:D0-coupling}
\end{equation}
The matrix operator in this expression which couples to the background
field is simply the dipole moment of the D2-brane charge encoded in
the system of D0-branes \cite{Dan-Wati-2,Mark-Wati-4}.  It was pointed
out by Myers in \cite{Myers-dielectric} that in the presence of such a
background flux, the lowest energy configuration for a system of
multiple D0-branes is a spherical membrane configuration
\cite{Dan-Wati} in which the noncommuting matrices $X^i$ are
proportional to $N$-dimensional generators of $SU(2)$.  If such a
membrane is placed in a nontrivial external B field, according to the
mechanism described in this note there should be additional
contributions to the D-particle number given by the integral of the B
field over the membrane world volume and from the bulk.  From
(\ref{eq:D0-coupling}) and the corresponding couplings between the
higher moments of the membrane charge and the background described in
\cite{Mark-Wati-4} it is clear that the spherical system of D0-branes
correctly acts as a source for the 4-form field, so that there will
indeed be an additional bulk contribution to the total D0-brane
charge.  However, since we know that the net D0-brane charge is really
$N$, there must be an additional term analogous to
(\ref{eq:IIA-Poynting}) in the nonabelian D0-brane action.  In
\cite{Mark-Wati-4}, the complete set of linear couplings of a system
of multiple D0-branes to supergravity background fields were deduced
from matrix theory.  The extra term we need here, however, will be a
term which couples quadratically to the supergravity background
fields.  By T-dualizing the 9-brane action as discussed in
\cite{Mark-Wati-5,Myers-dielectric}, it is possible to see that such a
term indeed appears.  Thus, we can see that in the language of
D0-branes the results of this note are again reproduced, and that $N$
is indeed the complete D0-brane charge.  An example of a situation in
which 4-form flux in M-theory is used both to blow up a graviton and
to induce (angular) momentum in the resulting membrane is discussed in
\cite{mst}.  Although in this case there is no circle on which
M-theory is compactified it would be interesting to understand this
picture better, possibly using the mechanism we have discussed here.

In this note we have focused on D2-branes.  It is natural to extend
this analysis to higher-dimensional D$p$-branes.  In general, on a
D$2p$-brane there is a coupling of the form $\int C^{(1)} \wedge B
\wedge F^{p-1}$.  The argument described here extends very easily to
this case.  We know that $F^{p-1}$ on a D$2p$-brane corresponds to
D2-brane charge, and the coupling in question thus describes precisely
the sort of 11-momentum discussed in this note for this D2-brane
charge.  On any D$p$-brane there is also a coupling of the form $\int
C^{(p-1)} \wedge B$.  These terms are very similar to those discussed
here, but do not have the physical interpretation in terms of M-theory
momentum; an example of a configuration in which a term of this type
is relevant appears in \cite{Polchinski-Strassler}.  There are also
Wess-Zumino terms in the world-volume action of a D$p$-brane which are
of quadratic or higher order in the B field.  The simplest example of
such a term is the term $\int C^{(1)} \wedge B \wedge B$ in the D4-brane
action.  It would be interesting to see whether terms of this form
have a simple interpretation analogous to the discussion in this note.

In this note we have discussed D2-branes which are wrapped on
homotopically trivial cycles in space-time.  When branes are wrapped
on nontrivial cycles, the physics can be more complicated.  For
example, when D2-branes are wrapped on a nontrivial space-time torus,
then the presence of a B field can be interpreted in terms of a
world-volume Yang-Mills theory on a noncommutative torus \cite{cds}.
Recently there have also been discussions of noncommutative geometry
in the context of spherical D2-branes in B fields
\cite{ars,Li-fuzzy,Pawelczyk}.  It would be interesting to understand
better how the discussion in this note fits into the framework of
noncommutative geometry produced by B fields.

\section*{Acknowledgements}

I would like to thank   S.\ Das, M.\ Douglas, I.\ Ellwood, J.\
Polchinski and J.\  Troost for helpful discussions and correspondence.
This work was supported in part by the A.\ P.\ Sloan Foundation
and in part by the DOE through contract \#DE-FC02-94ER40818.

\normalsize
\newpage

\bibliographystyle{plain}

\end{document}